# An End-to-End Explainable AI Framework with Automated LLM-Based Natural Language Explanation Generation for Energy Systems

**Venkata Sesha Sai Raj Nanduri[1], Akthar Hussain[2] and Van-Hai Bui [1,*]**

[1] Department of Computer and Information Science, University of Michigan-Dearborn, 4901 Evergreen Rd, Dearborn, MI 48128, USA

[2] Department of Electrical and Computer Engineering, Laval University, 2325 Rue de l'Université, Québec, QC G1V 0A6, Canada

**Abstract**

Explainable AI (XAI) is important for deploying machine learning systems in domains where stakes are very high and where transparency, trust and accountability are critical. Although black box models like deep neural networks often perform with high efficiency, interpreting their decisions remains a difficult task. This paper proposes a reusable end-to-end XAI framework that is the combination of prediction, explanation generation, evaluation and converting these explanations into natural language text of explanation which can be easily understood by the non-technical stakeholders as well. This framework initially preprocesses tabular datasets using standardized feature transformation techniques and trains deep neural network for both classification and regression tasks. Local and global explanations are generated using XAI algorithms like Local Interpretable Model-agnostic Explanation (LIME) and SHapley Additive exPlanations (SHAP) respectively. To evaluate these explanations, we use metrics like fidelity and stability to know how accurately and consistently explanations reflect the model behavior. The generated explanation includes feature importance scores, prediction specific attributes. These are then transformed into a structured input to the Large Language Model (LLM), which generates a natural language explanation through which everyone can understand the explanations generated by XAI algorithms. This framework is tested on power system fault dataset detection dataset and building energy labels dataset. The experimented results demonstrate the effectiveness of the proposed approach across both classification and regression tasks. For fault detection, the neural network model achieved 99% accuracy with ROC-AUC score of 1.00. For building energy prediction, model achieves $R^2$ score of 0.67. Furthermore, fidelity and stability also stays at the top with a nearly perfect score which makes the model very reliable. These findings says that the proposed approach produces a stable and faithful explanations while improving the interpretability of black box model to everyone with the help of LLMs. This framework provides a model agnostic solution for enhancing transparency in both classification and regression tasks.

## 1. Introduction

Artificial Intelligence (AI) and Machine Learning (ML) have become the integral part across numerous application domains, including healthcare, finance, transportation and energy systems etc. Their ability to automatically learn the complex patterns from huge datasets has enabled significant improvements in performance and decision-making. However, as AI systems are most frequently used in the high-stakes domains where decisions directly affect the lives of people, the ability to understand and justify their predictions has become equally important. This growing demand has led to the rapid development of Explainable Artificial Intelligence (XAI), which aims to make machine learning models more transparent, interpretable and trustworthy while preserving the performance [1 – 3].

Among the ML techniques, deep neural networks are good at solving difficult problems because of their ability to learn highly nonlinear relationships from huge amount of data. Consequently, they have been widely adopted for applications such as power system fault diagnosis, computer vision, natural language processing and energy prediction where accurate and interpretable predictions are essential for ensuring trust and sustainable resource management. Though, they are good at their predictions, these operate as a black box model i.e. they do not provide insights about their decision. The internal representations learned by neural networks are often too complex to understand which makes it difficult to figure out which input features influence the decision made by neural network. This lack of transparency is the biggest obstacle to use black box models in the high-stakes applications, where understanding the reason behind predictions is as important as the predictions themselves. Therefore, improving interpretability of black box models has become an important research direction in XAI [4 – 6].

To improve model transparency, several post-hoc explanation techniques have been proposed. One of the earliest and most widely used approaches is Local Interpretable Model-agnostic Explanation (LIME), which explains local predictions by approximating the complex model with a local surrogate model [7]. Since LIME is model agnostic, it can be applied to both classification and regression tasks. Another widely used approach is SHapley Additive exPlanations (SHAP), which employs Shapley values from cooperative game theory to quantify the contribution of each feature towards a prediction [8]. Unlike LIME, SHAP can provide explanations for both local aa well as global explanations describing the overall behavior of a model. Further studies have extended SHAP to efficiently interpret tree-based models and improve global model understanding [9].

Beyond feature attribution methods, several alternative explainability techniques have been proposed to improve model interpretation from different perspectives. Ribeiro et al. introduced Anchors, a rule-based explanation framework that generates high-precision decision rules for individual predictions [10]. Similarly, another technique is Testing with Concept Activation Vectors (TCAV) explains neural network decisions using human understandable concepts rather than individual input features [11]. Surveys and interpretability frameworks further classified inherently interpretable models and post-hoc explanation techniques while highlighting their respective strengths and weaknesses [12 – 14]. These surveys tells us that no explanation technique is applicable universally and combining multiple explanation methods provides more understanding of complex ML models.

Although decent progress has been made in developing explainability techniques, evaluating them remains the major challenge so that we can confirm whether we can trust the model's interpretation. Previous studies have shown that many explanation methods may produce unstable or inconsistent results under slight input variations, which enhances the need of evaluation metrics [15]. When we talk about evaluation metrics fidelity and stability have emerged as a crucial measure for assessing explanation quality by exactly evaluating how accurate the explanations are and how consistent are they [16].

Recent advances in Large Language Models (LLMs) have further expanded the capabilities of XAI by making these interpretations more meaningful even to non-technical stakeholders by converting these explanations into natural language explanations. In other words, to interpret the numerical attributions or plots generated from XAI algorithms like LIME and SHAP often require domain expertise. But we want our interpretations to be understood by everyone. So, to bridge this gap we take the help of LLMs. Slack et al. demonstrated this concept through TalkToModel, which allows users to interact with ML models using natural language [17]. Similarly, there are few developments in prompting and instruction tuning through Reinforcement Learning (RL) which improved reasoning and explanation capabilities of large language models [18 – 20]. Beyond general-purpose AI applications XAI nowadays has also gained importance in sustainable energy systems, based on recent studies XAI techniques improved transparency in building energy management and electricity consumption, demonstrating that explainability can enhance user trust and facilitate energy efficient building management [21 – 23].

Despite these advances, several research gaps remain. Most existing studies either focus on local explainability like LIME or global explainability like SHAP, without trying out both the techniques in a unified framework. Moreover, many studies emphasize on visualization plots while providing limited quantitative evaluation of explanation techniques using metrics like fidelity and stability. Existing explainable AI applications sustainable energy systems have similarly focused on specific prediction tasks, such as building energy forecasting, rather than developing reusable frameworks that can be adapted across multiple domains [21 – 23].

Motivated by these challenges, this journal proposes a unified explainable AI framework for interpreting neural network models across both classification and regression tasks. The framework further combines LIME and SHAP to generate local and global explanations, evaluates explanation quality using fidelity and stability metrics, and employs a LLM to automatically generate the natural language explanations to LIME and SHAP results. The framework is validated-on power system fault detection and building energy prediction datasets.

The major contributions of this work are summarized as follows:

- A reusable end to end XAI framework is proposed for neural network-based classification and regression tasks.
- The framework integrates LIME and SHAP to provide local and global explanations.
- Explanation quality is evaluated using the metrics like stability and fidelity to improve reliability of model
- A LLM is incorporated to automatically convert technical explanation into concise and human-readable explanations.
- The proposed framework is validated on power system fault detection dataset, and a building energy labels dataset, demonstrating that it can produce accurate predictions together with faithful, stable and interpretable explanations.

## 2. Methodology

This section describes the end-to-end XAI framework, detailing each stage of the pipeline from data-processing to explanation generation in natural language using LLMs which can be easily understood by all the users. This methodology is designed to be dataset-agnostic, modular and applicable to both classification and regression tasks.

The proposed workflow follows a structured pipeline in which raw tabular data is transformed into predictions using black-box models, followed by post-hoc explanation generation, quantitative evaluation of explanations, and finally translation of the prediction using LLM into a human interpretable explanation.

This framework operates on various datasets which contains both numerical and categorical features. Prior to model training all these datasets undergo a standard preprocessing procedure to ensure consistency and prevent data leakage.

Numerical features are standardized using z-score normalization, while categorical features are encoded using one-hot encoding. A unified preprocessing pipeline is implemented using column transformer which applies same transformations to be applied consistently during both training and inference. All preprocessing steps are encapsulated within a pipeline to ensure reproducibility and facilitate seamless dataset replacement.

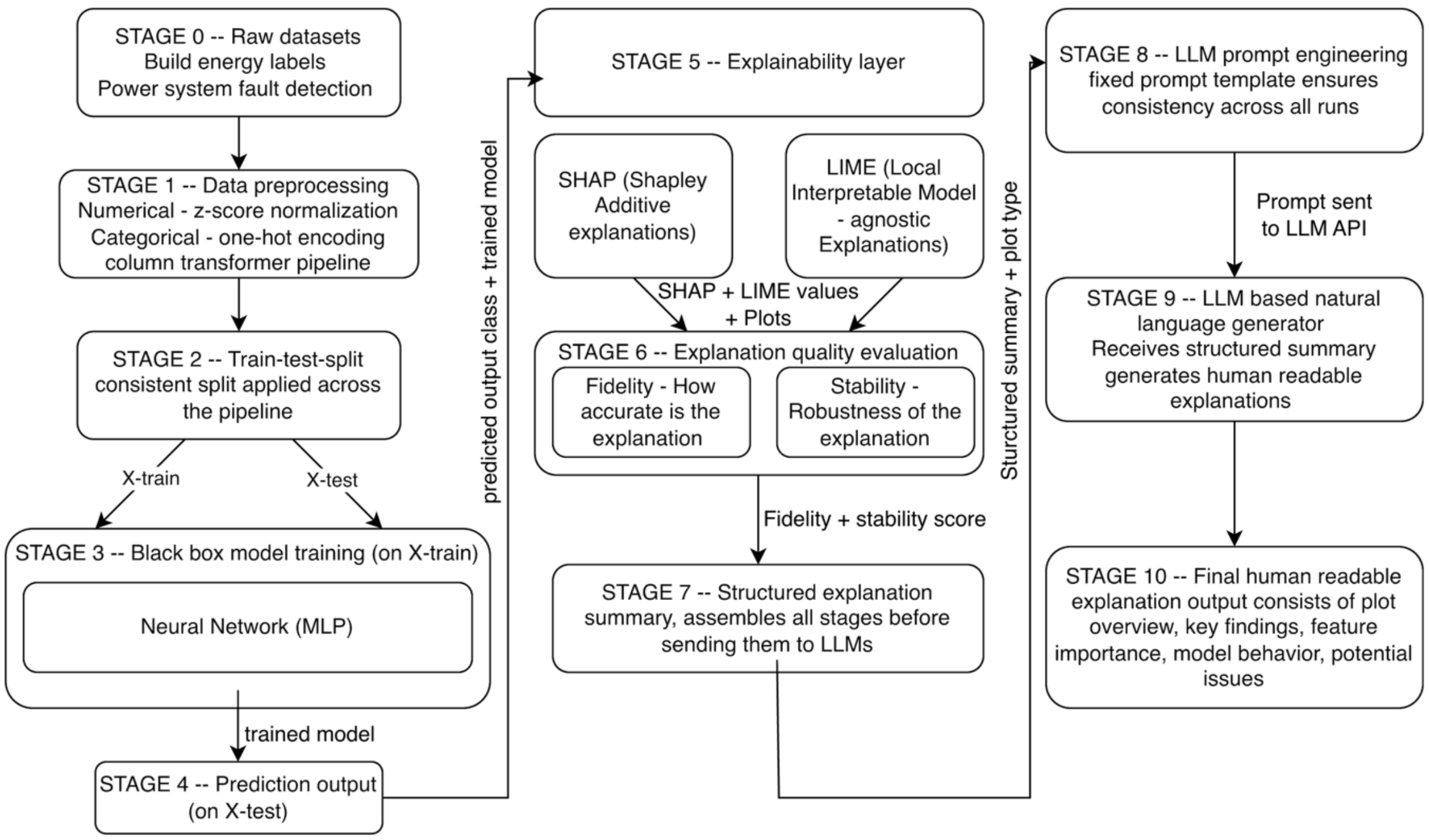

a. Black box prediction workflow. b. Explainability workflow. c. LLM Integration workflow

Fig 1 : The flowchart of the Explainable AI framework

Fig 1a illustrates the black box prediction workflow, this starts with collection of the datasets, including building energy labels and power system fault detection. The first stage includes steps like data preprocessing in which numerical features are standardized using z-score normalization and categorical features are transformed using one-hot encoding. These preprocessing steps are implemented using column transformer pipeline, ensures that the transformations are applied consistently while preventing data leakage between the training and testing datasets. After preprocessing, the dataset is divided into training and testing datasets with the help of train-test-split, with training data (X_train) used for model learning and test data (X_test) reserved for evaluation. The black-box model is then trained using a Multilayer Perceptron (MLP) neural network, which consists of two hidden layers for classifier containing 192 neurons with the ReLU activation function and three hidden layers for regressor containing 448 neurons with ReLU activation function. The network is optimized using the adam, with L2 regularization (alpha = 0.0001 for classifier & 0.00001 for regressor) to reduce overfitting and a maximum of 500 training iterations for classifier and 1000 iterations for regressor. Once the model has been trained, it is applied to the unseen test data to generate prediction outputs, which are subsequently used for performance evaluation and explainability analysis.

Fig 1b tells us about the explainability workflow this comes straight after the black box model workflow. The process begins with passing of trained neural network model and its prediction to the explainability layer, where Local Interpretable Model agnostic Explanation (LIME) generates local explanations for individual predictions by approximating the model with a small surrogate model around a particular instance, whereas Shapley Additive explanations helps us to explain model globally as well as locally by assigning some importance score to the features which tells us which feature is crucial for the decision made by model. These results are then evaluated using the evaluation metrics namely explanation stability which assesses the consistency of the explanation and fidelity measures how well the explanation reflects the original model's behavior. Overall, the workflow compiles model information, prediction, important features, direction of feature influence, and explanation quality combines and form a structured summary.

Fig 1c explains the LLM integration workflow it is responsible for generating human readable explanations from the structured outputs of the explainability framework. It begins with the structured explanation summary and the selected plot type (SHAP, LIME) which are used to construct a standard prompt template. This prompt is then sent to the Large Language Model (LLM) which generates the natural language explanations using the structured information from the previous workflow. This explanation consists of 5 points which mainly focuses on the aspects like plot overview, key findings, most influential features, the model's decision-making behavior, and any potential issues such as biases or limitations, making the results accessible to both technical and non-technical stakeholders.

**2.1 Model Training**

The neural network implemented in this framework is a Multi-Layer Perceptron (MLP). This is done using MLPClassifier and MLPRegressor from scikit-learn. The input layer consists of d number of neurons (d = 6 for fault detection, d = 7 for energy regression). There are 2 hidden layers for MLPClassfier consists of 192 neurons and 3 hidden layers for MLPRegressor consists of 448 neurons both with ReLU activation function. The output layer completely depends on the type of neural network we use.

Now we will have a look at the feed forward pass, when the neural network gets an input vector x $\in R^d$, it does two things first it calculates the pre activation of each hidden neuron j is computed as the weighted sum of inputs plus a bias:

$$Zj = \sum_{i=1^d} Wij * Xi + bj \tag{1}$$

where $w_{ij}$ is the weight connecting input neuron i to hidden neuron j, and $b_j$ is the bias of hidden neuron j. The tanh activation function is then applied to produce the hidden layer output:

$$aj = ReLU(Zj) = max(0, Zj) \tag{2}$$

The ReLU activation function is used because it outputs the input directly when its positive and zero otherwise. Enabling the efficient gradient propagation while reducing the vanishing gradient problem which is commonly encountered in neural networks. Also, ReLU introduces computational simplicity by requiring only a threshold operation, leading to faster training and improved convergence on standardized input features. The output layer then produces the final prediction. For fault detection task, softmax is applied to generate class probabilities, whereas for the energy regression task, a linear activation produces the log-transformed energy value directly

Now let's look at the loss function and L2 regularization, for the fault detection classification task, we use cross-entropy loss over all n training samples. For the energy regression task, we used mean squared error loss. In both cases L2 regularization is added to the loss to penalize large weights and prevent overfitting. The regularized objective is:

$$Lreg = L(\Theta) + (\alpha/2)\,\|\Theta\|^2 \tag{3}$$

Where θ represents trainable weight and bias parameters of the network, L(θ) is the task-specific loss. The L2 penalty shrinks all weights toward zero proportionally to their magnitude, discouraging the model from assigning excessive importance to any single feature – a particularly important property given that the fault detection input features have very large dynamic ranges even after z-score normalization.

Looking at the backpropagation and optimization, basically the model parameters are learned using backpropagation, which applies the chain rule of calculus to compute the gradient of the regularized loss with respect to every weight and bias in the network. For the output layer weights, the gradient is:

$$\partial Lreg/\partial W(2) = \delta(2)(a)^T + \alpha W(2) \tag{4}$$

Where $\delta^{(2)}$ is the output layer error signal. The error is now propagated back through the ReLU activation to compute the hidden layer gradient:

$$\delta_j(1) = (W(2)^T\delta(2))_j.ReLU'(Zj) \tag{5}$$

Here, ReLU′(Zj) is the derivative of the ReLU activation function, which equals 1 when $Zj > 0$ and 0 otherwise. This derivative allows gradient to propagate only through activated neurons, thereby reducing the vanishing gradient problem and enabling efficient learning. The weights of the input layer are then updated, the network parameters are optimized using adam optimizer, which combines momentum with adaptive learning rates by maintaining exponentially decaying estimates of the first and second moments of the gradients. This results in faster convergence, improved optimization stability, and effective training of the neural network for both the fault detection and energy regression tasks.

The evaluation metrics we use for this are: Accuracy, F1-score, and ROC-AUC for fault detection classification task; $R^2$, Mean-squared error (MSE), for the energy regression task. This evaluation protocol enables direct and fair comparison between the two model architectures across both tasks.

### 2.2 Explainability Layer

Post-hoc explainability methods are applied to explain the predictions generated by the block-box model. This layer operate independent of training process and does not interfere with model parameters.

#### SHAP-Based Explanations

SHAP, which stands for SHapley Additive exPlanations is a way to explain what is going on with a model after it has made a prediction. It looks at each feature that affects the model prediction and gives it a score. This score is called a Shapley value. It shows how much each feature influence the prediction.

Think of it like a game where each feature is a player and the prediction is the prize. The prize must be divided among all the players. The shapley value $\varphi_i$ for feature i is defined as the weighted average of its marginal contributions across all possible subsets S of feature set F:

$$\psi_i(f,x) = \sum_{S\subseteq F(i)} [|S|!(|F|-|S|-1)!/|F|!] * [f(S\cup i) - f(S)]$$

(6)

Where F is the set of features and S is a subset of features excluding feature i, f(S) denotes the expected model output when only the features in S are known (all others marginalized out), and the binomial coefficient $|S|!\ (|F| - |S| - 1)!\ /\ |F|!$ weights each subset by the probability that features arrive in an order where S precedes i. The shapley values satisfy three key properties: efficiency, symmetry and consistency.

**KernelSHAP**

For the neural network, TreeSHAP is not applicable. Instead, KernelSHAP is used, so it is a way to estimate the importance of each feature in a model. It does this by looking at how different features work.

To do this for an instance KernelSHAP creates many random groups of features. For each feature it decides whether to include it or not. If a feature is included it is like the feature is there. If it is not included it is like the feature is not being used.

KernelSHAP then makes a prediction for each of these groups. It does this by replacing the features that are not being used with values, from the same kind of data. After that it solves a math problem to figure out how important each feature is. KernelSHAP is a way to do this because it can be used with any kind of model not just neural networks.

$$\psi^* = argmin_\psi \sum_{m=1}^{M} \Pi(z'm)[f(h(z'm)) - g(z'm)]^2$$

(7)

where $g(z') = \varphi_0 + \sum\varphi_i z'_I$ is the linear explanation model , h(z'm) maps the binary coalition back to the original feature space, and $\pi(z')$ is the SHAP kernel weight:

$$\Pi(z') = (|F| - 1)/[C(|F|, |z'|) * |z'| * (|F| - |z'|)]$$

(8)

This kernel weight gives high importance to coalitions with very few or many features, which correspond to the most informative marginal contributions. The solution $\varphi^*$ to this weighted regression yields approximate shapley values that satisfy local accuracy: $\sum\varphi_i + \varphi_0 = f(x)$, where $\varphi_0 = E[f(X)]$ is the model's baseline expected output.

**SHAP Evaluation Metrics**

The quality of SHAP explanations is looked at using two measures: fidelity and stability. Both these measures use the way SHAP explanations are broken down into parts.

Fidelity for SHAP explanations uses the fact that SHAP's very efficient. Since SHAP makes sure that the sum of all the parts equals the value minus the average value. For KernelSHAP the weighted regression is an approximation, so fidelity quantifies how close the approximation is. In this project, fidelity is computed as:

$$Fidelity_{SHAP} = 1 - MSE(f(X), \psi_0 + \sum_i \psi_i(X))/Var(f(X))$$

(9)

Stability measures how sensitive the shapley value vector is to small perturbations in the input. For instance, x a perturbed copy is generated and SHAP values are recomputed. The stability is:

$$Stability_{SHAP} = 1 - E[||\psi(x) - \psi(x')||_2]/E[||\psi(x)||_2] \tag{10}$$

The fault detection models are stable because they have a clear decision boundary. The energy neural network is not as stable because its SHAP values are highly sensitive, to small changes.

**LIME-Based Explanations**

LIME system helps us understand how the computer makes predictions. It does this by creating a model that is easy to understand, which we will call the surrogate model. The surrogate is constrained to be interpretable and is fitted only in the vicinity of x, not globally.

**LIME objective function:**

The explanation g* is found by solving the following optimization problems:

$$g^* = argmin_{g \in G} L(f, g, \Pi_x) + \Omega(g) \tag{11}$$

Where G is the class of interpretable models, L is locality weighted loss measuring how well g approximates f near x, $\pi_x$ is the proximity measure between x and perturbed samples and Ω(g) is a complexity penalty on g. In this project, Ω(g) is kept low by restricting LIME to report the top-K most influential features per instance.

**Locality weighted Loss proximity Kernel:**

The locality weighted loss L measures the fidelity of the surrogate model to the black box model over perturbed samples z drawn from $\pi_x$:

$$L(f, g, \Pi_x) = \sum_{z \in Z} \Pi_x(z)[f(z) - g(z)]^2 \tag{12}$$

The proximity weight $\pi_x(z)$ is computed using an exponential kernel on the euclidean distance between the perturbed sample z and target instance x:

$$\Pi_x(z) = exp(-d(x, z)^2/\sigma^2) \tag{13}$$

where d(x, z) is the euclidean distance and σ is a bandwidth parameter controlling the size of local neighborhood. Perturbed samples close to x receive high weights and heavily influence the surrogate fit, while distant samples receive near zero weights and are effectively ignored. This ensures that the surrogate g captures local behavior of the black box model f around the specific instance being explained rather than its global behavior.

**Linear surrogate and Feature coefficients**

The surrogate model fitted by LIME is a weighted linear model:

$$g(z) = w_0 + \sum_{i=1}^{|F|} w_i z_i$$

(14)

where $w_i$ is the coefficient of feature i in the surrogate. These coefficients are the LIME feature importances displayed in the bar plots in the result section. When we look at the feature importance a positive weight means the feature makes the prediction higher than the average. This is shown in green. On the other hand, a negative weight means the feature makes the prediction lower than the average, this is shown in red.

In this project, for the fault detection task the surrogate uses logistic regression (classification), while for the energy regression task it uses weighted ridge regression – both solved using the perturbed samples Z = {z1, ……, zN} (N = 5000 samples by default) drawn by randomly masking features of x with values sampled from their training distribution.

**LIME Evaluation Metrics in this project**

LIME is different from SHAP because LIME does not make sure it works efficiently. So, fidelity is evaluated differently. For LIME, fidelity measures how well the locally fitted surrogate g reproduces the black box model f on the perturbed neighborhood of each test instance. The local $R^2$ of the surrogate on the neighborhood samples serves as per-instance fidelity measure:

$$Fidelity_{LIME}(x) = 1 - [\sum_z \Pi_x(z)(f(z) - g(z))^2 / \sum_z \Pi_x(z)(f(z) - f)^2]$$

(15)

Where f is the average of f(z) over the neighborhood. A high local $R^2$ shows that the surrogate g closely follows the model's predictions near x which means the LIME coefficient wi are good representations of how the model behaves locally. In this project we measure LIME fidelity by averaging the $R^2$ across all test cases. In this task of detecting faults LIME works well for the neural networks because features can be easily separated in a way near the decision boundary.

In the regression LIME does not work well because the relationship between building type indicators and energy use is very non-linear, in the neural networks model.

LIME stability is evaluated by repeating the explanation for the same instance x multiple times (K = 10 repetitions) with different random perturbation seeds and measuring the variance of the resulting coefficient vectors

$$Stability_{LIME}(x) = 1 - (1/K) \sum_{k=1}^{K} ||w^k(x) - \bar{w}(x)||_2 / ||\bar{w}(x)||_2$$

(16)

The coefficient vector from the k-th repetition is called w(k)(x). The mean coefficient vector across all k repetitions is called w¯(x). When the stability is close to 1.0 it means that LIME gives the same feature attributions every time no matter which random perturbation samples are used. This shows that the surrogate has found a local linear approximation that works well.

LIME has a problem: it uses sampling of the neighborhood so its explanations can be different each time it is run even for the same instance. This happens especially when the decision boundary of the black-box model is not linear or is irregular. In this project the LIME explanations for the fault detection task are very consistent across repetitions. This is because both models have smooth and near-linear decision boundaries in most of the feature space. However, the LIME explanations for energy regression are more variable for instances near building type boundaries, in the neural network.

**2.3 LLM-Based Natural language Explanation Generation**

- Motivation for LLM Integration

  Traditional explainability methods gives us insights through numerical feature attributes and graphical visualizations. While these outputs are very informative, but they cannot be interpreted by all the users. For example, a SHAP summary plot may show that a feature contributes positively or negatively to the prediction but interpreting the overall implication of these contributions requires domain and technical knowledge.

  LLM have recently demonstrated strong capabilities in natural language understanding and generation which enables the structured information about the prediction in a very understandable way to each and every user.

- Input representation to the LLM

  To generate natural language explanations, the LLM is not directly given raw plot images. Instead, structured information extracted from SHAP and LIME explanations is provided as input. The structured input contains the following components:

  1. Model information: The name and type of which black box used (Neural network).
  2. Prediction Output: The predicted class label for classification tasks or the predicted numerical value for regression tasks.
  3. Top contributing features: A ranked list of features with the highest SHAP values indicating their contribution magnitude.
  4. Direction of feature influence: Whether the feature influence prediction positively or negatively.
  5. Reliability metrics: Fidelity and stability scores which is used to measure the reliability of model.

- Prompt engineering

  For the maintenance of consistency and variability a fixed prompt template is necessary. Prompt engineering plays a crucial role in controlling the structure and length of generated natural language explanation.

  The prompt used for this framework is:

```
prompt = f"""Analyze this {plot_type} explainability plot from a machine learning model and provide:

    1. **Plot Overview**: Brief description of what the plot shows
    2. **Key Findings**: Main insights and patterns visible in the plot
    3. **Feature Importance**: Which features are most important/influential (if applicable)
    4. **Model Behavior**: This reveals about how the model makes decisions
    5. **Potential Issues**: Any concerning patterns or biases to watch for
```

Explain all these points clearly and very concisely in about 5 bulleted sentences.
{f'Additional Context: {context}' if context else ''}
Be concise but thorough. Use clear language suitable for non-technical stakeholders."""

- Output of the LLM explanation Layer
  The final output produced by the LLM is the short explanation which contains of the 5 bulleted points which contains plot overview, key findings, feature importance, model behavior and potential issues which explains the output predicted by the explainable models. For example, the generated LIME explanation for Neural networks.

**2.4 End-to-End pipeline integration**

All components are integrated into a unified pipeline that enables seamless end-to-end execution. Once the dataset is provided, the system end-to-end execution. Once the dataset is provided, the system automatically performs preprocessing, model training, prediction, explanation generation, evaluation and LLM output generation.

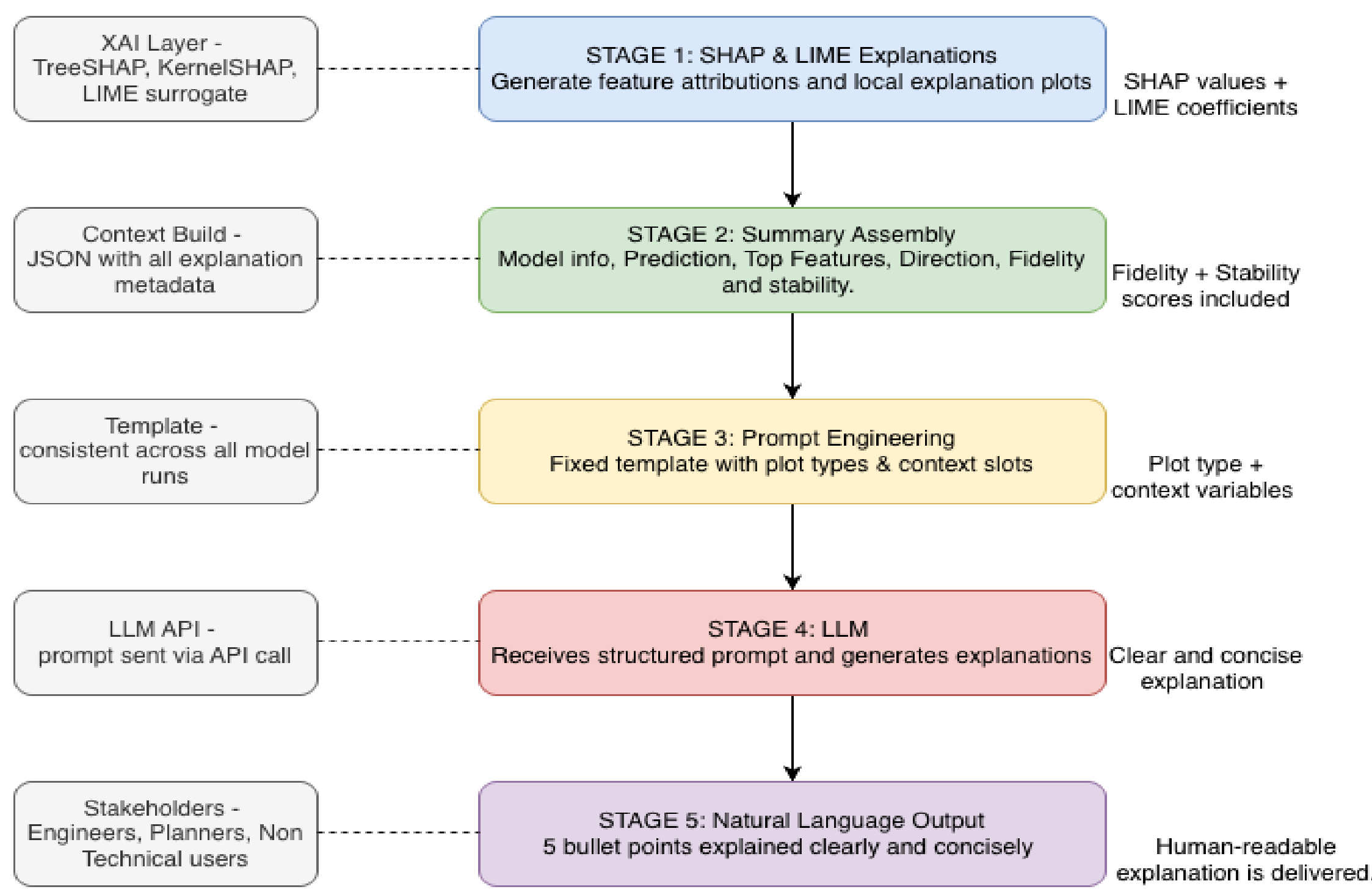

Fig 4: LLM Integration

## 3. Datasets

To evaluate the robustness of the proposed XAI framework, so the experiments were conducted on multiple benchmark datasets for both classification and regression tasks. The following datasets which contains mixture of numerical features and categorical features helps us to check the robustness of this framework.
These Datasets were selected to evaluate the proposed framework across diverse domains, ensuring that explainability methods are robust to variations in feature types, dimensionality, and prediction tasks.

- Build energy labels dataset

  The building energy labels dataset evaluates the proposed framework on a regression task in the domain of urban energy management and sustainability. This dataset is employed to assess the framework's capacity to generate faithful and stable explanations for continuous numerical predictions.

- Power system fault detection dataset

  The Power system fault detection dataset evaluates the framework on a binary classification task in the domain of electrical energy. It is employed to examine the ability of proposed framework to produce reliable and interpretable explanations for safety-critical fault detection decisions.

### 3.1 Building energy labels dataset

This dataset consists of 2503 records of residential, commercial, educational and institutional buildings, collected as part of a municipal energy benchmarking initiative. Each datapoint contains structural and operational attributes of a building alongside its measured energy consumption and estimated carbon footprint. The dataset is clean with no missing values across all ten features, making it directly suitable for model training without imputation.

The primary regression target is site energy use (kBtu), which tells us the total energy consumed at a building site in thousand British thermal units. The predictor features include Year built, building age, number of floors, total gross floor Area (PropertyGFATotal), and building type, which is a categorical variable encoding four classes: Residential (58.4%), Commercial (35.2%), Educational (4.0%), and Institutional (2.3%). Building age is derived directly from year built and therefore only one of these two features is retained during training to avoid multicollinearity.

This dataset is particularly suited for evaluating the proposed XAI framework in a regression context because energy consumption is an outcome shaped by multiple interacting factors, including building age, size, type and occupancy patterns. Explaining which features drive predicted energy use is of direct practical value to urban planners and policy makers seeking to prioritize energy efficiency interventions.

### 3.2 Power system fault detection dataset

The Power system fault detection dataset contains 12,001 records of three-phase electrical measurements captured from a power transmission or distribution system. Each record consists of 6 real valued sensor readings, namely the current measurements Ia, Ib and Ic (in amperes) and the phase voltage measurements Va, Vb and Vc, along with a binary target label output that indicates whether the system is operating under a normal condition (0) or experiencing a fault (1). The dataset has no missing values across all meaningful columns.

The class distribution is also well balanced, with 6505 normal instances (54.2%) and 5496 fault instances (45.8%), removing the need of resampling techniques. The three phase current readings exhibit large dynamic ranges with standard deviations exceeding 350 A, as fault conditions typically induce significant current imbalances across phases. In contrast, voltage readings remain tightly bounded within approximately +/- 0.66 per unit, consistent with per-unit normalization conventions used in power system analysis.

This dataset is appropriate for evaluating the XAI framework in a safety-critical classification setting. Fault detection in power systems demands not only high predictive accuracy but also transparent and interpretable reasoning, as incorrect predictions can have severe consequences for grid safety. Applying SHAP & LIME to explain individual fault predictions enables domain engineers to verify whether the model is attending to physically meaningful signal patterns, thereby building trust in the automated detection systems.

| Attribute | Building Energy Labels | Power systems fault detection |
|---|---|---|
| **Task Type** | Regression | Binary Classification |
| **Domain** | Urban Energy Management | Electrical Power systems |
| **No of Records** | 2,503 | 12,001 |
| **No of Features** | 7 (4 numerical, 1 categorical with 4 classes) | 6 (all numerical: 3 currents, 3 voltages) |
| **Target Variable** | Site Energy Use (kBtu) – log transformed | Output (s): 0 = Normal, 1 = Fault |
| **Class/ Label distribution** | Residential 58.4%, Commercial 35.2%, Educational 4%, Institutional 2.3% | Normal 54.2%, Fault 45.8% - Balanced |
| **Missing values** | None | None |
| **Pre-processing applied** | Z – score normalization; One-hot encoding; log transform on target | Z – score normalization only (since all features are numerical) |
| **XAI purpose** | Explain energy predictions; Support urban planning decisions | Explain fault decisions; verify physical signal patterns |

Table 1: Summary of Datasets

## 4. Results and Observations

This section tells us all about the experimental results obtained from the XAI framework when tested with power system fault detection dataset using two black box predictors: neural network. If we look at the results we have performance metrics, LIME explanations, SHAP Explanations and explanation quality metrics.

#### 4.1 Fault detection

#### 4.1.1 Model performance

The performance of Neural network classifier was evaluated using accuracy, precision recall, F1-Score and ROC-AUC on both training and testing data. As shown in the figure below model achieved training accuracy of 0.995 and a testing accuracy of 0.990, which indicates marginal performance difference between the two datasets. Similarly, the training and testing F1-scores were same, demonstrating that the classifier maintained a balanced trade-off between precision and recall while generalizing effectively to unseen data. The ROC-AUC score of 1.000 on both datasets further confirms the model's excellent discriminative capability across different decision thresholds. The close agreement between the training and testing metrics suggests that the proposed neural network exhibits minimal overfitting while maintaining robust predictive performance. These results indicate that the classifier effectively captures the underlying relationships between electrical measurements and fault conditions, making it suitable for reliable fault detection in power system applications.

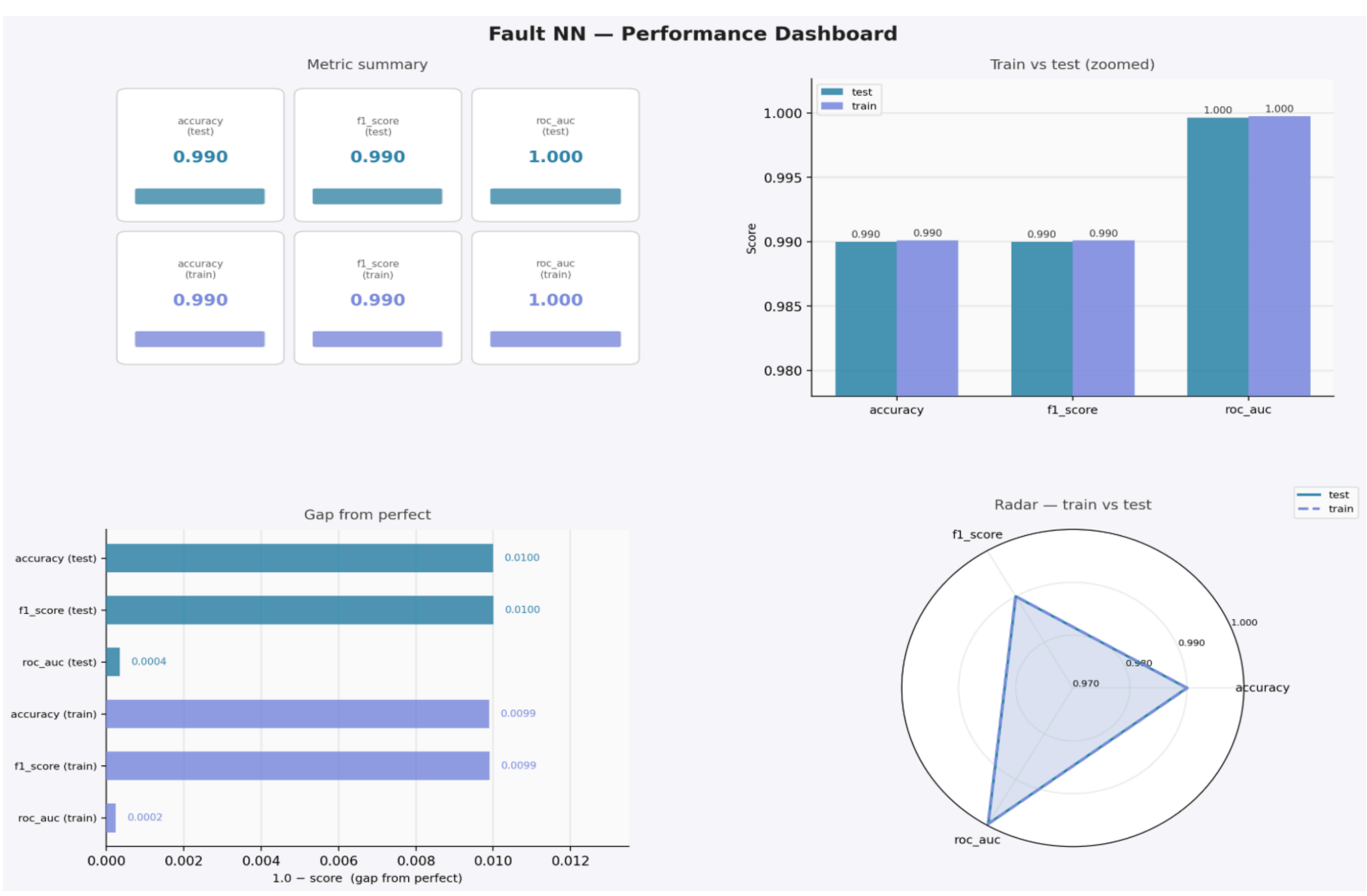

Fig 5: Fault detection performance metrics

#### 4.1.2 LIME Based local explanations

To improve the interpretability of the decisions made by neural network, LIME was used to generate local explanations for two representative test instances: one that was correctly classified and another that was misclassified. These local explanations shows us how individual input features contribute to the model's prediction and provide insight into the model's decision-making process. In the plots below left plot is misclassified and right one is correctly classified plot where positive feature contribution (green) support the predicted class, while negative contributions (red) oppose it.

For misclassified instance, current_c, voltage_a, voltage_b and current_a contribute positively towards predicted class, whereas voltage_c provides strongest negative contribution compared to current_b with smaller contribution. Since, negative contribution from voltage c is so dominant it outweighs the positively contributing features, which results in predicting incorrect class. This indicates that the model encounters difficulty when highly influential features providing negative evidence, resulting ambiguity near the decision boundary.

In contrast, when we look at the correctly classified plot it exhibits more coherent pattern of feature contributions. Although voltage_c, voltage_a contribute negatively with a very small negative contribution of current_b, they are outweighed by strong positive contribution of voltage_b and smaller contribution of current_ a and current_c together enabling the classifier to correctly identify the fault class.

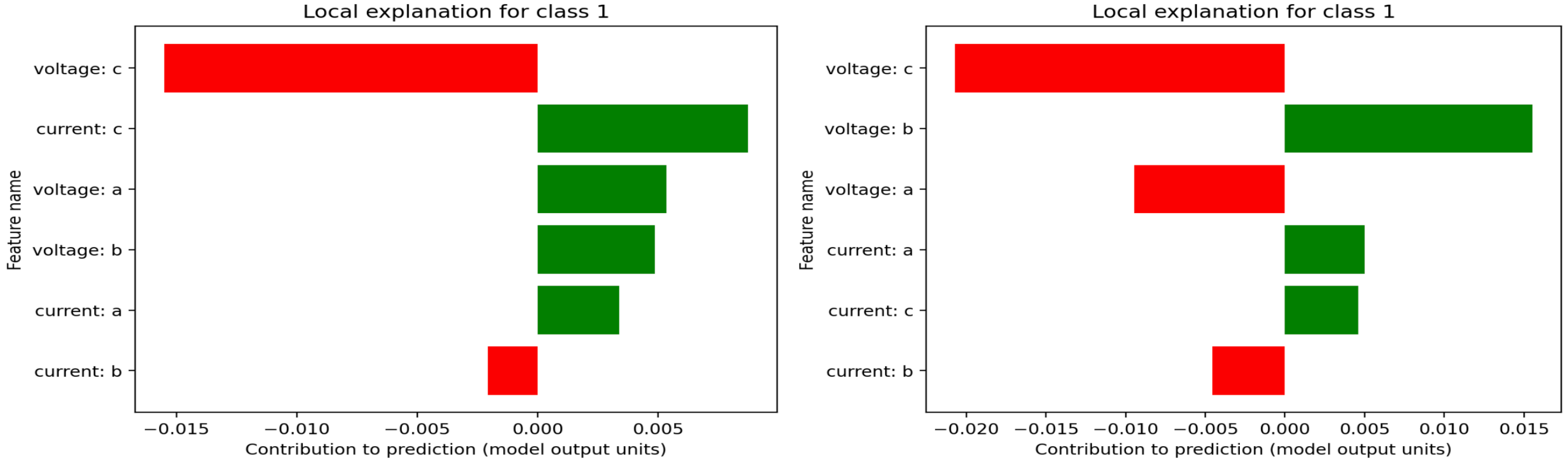

Fig 6: LIME explanations of 2 instance

**4.1.3 SHAP Based Global and Local Explanations**

Global feature importance was computed using KernelSHAP across the test set. The SHAP bar plot ranks all six features by their mean absolute SHAP value. The current in phase A (current a) is the most important thing that affects the result. It has an impact with mean SHAP value 0.13. The currents in phase B, (current b) and the voltage in phase C (current c) are also very important with a value of 0.10 each. The Voltage in phase B and phase A are important too with a value of 0.09, whereas the current in phase C has the least impact with a value of 0.07.

By looking at these results, we can say that neural network does not simply rely on single dominant feature, instead it learns information from all the features to characterize fault conditions. The model learns physically meaningful relations that align with domain knowledge while exploiting complementary information from multiple electrical signals.

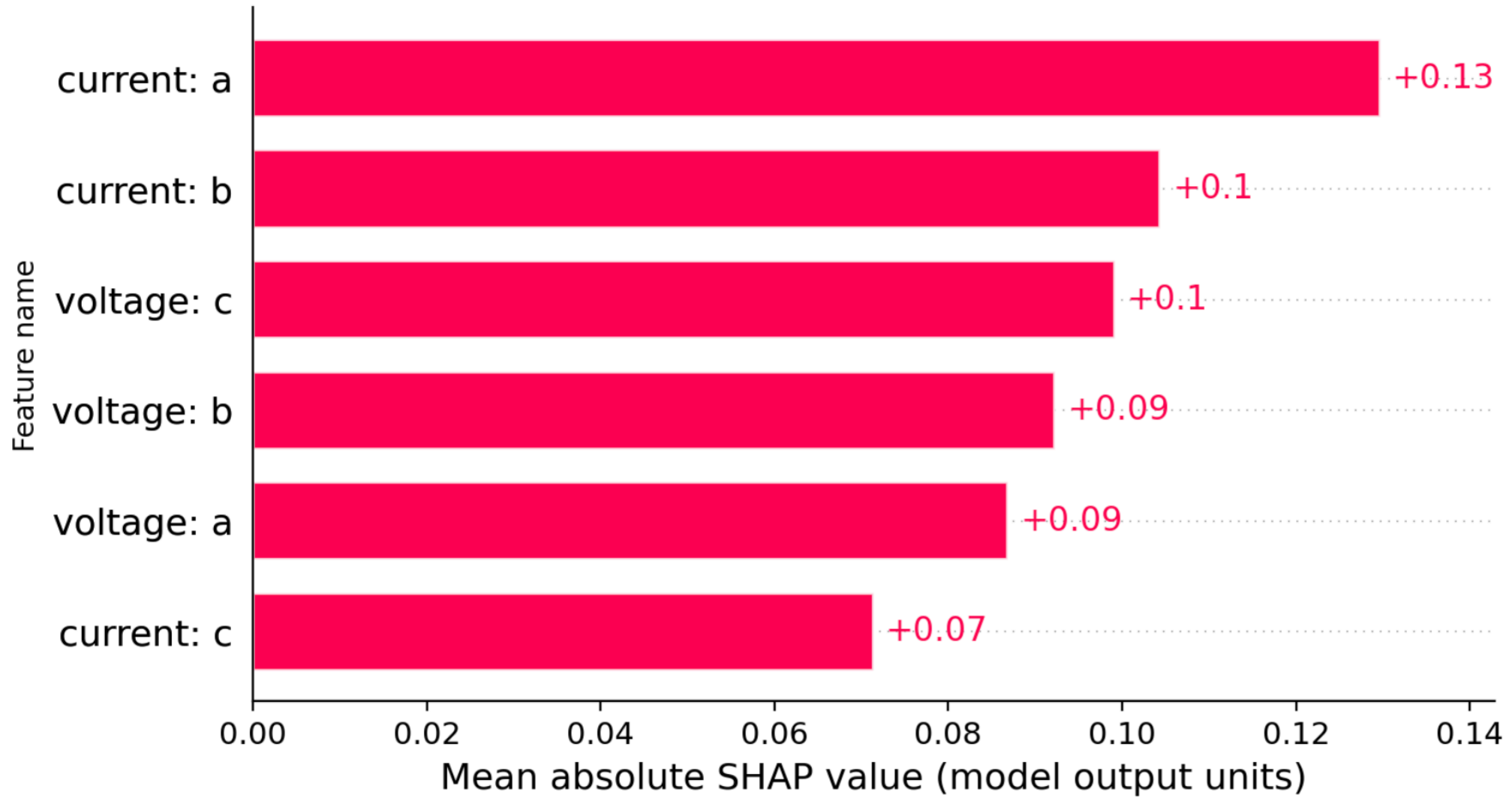

Fig 7: SHAP bar plot

The SHAP beeswarm summary plot helps us understand how each feature influence the result. The distribution of the SHAP values in the below plot demonstrates that contribution of each feature varies across different operating conditions, which reflects the nonlinear relationships that neural network learned. When we look at current_a we observe both positive and negative SHAP values depending on input which tells us its influence is dependent on the context. Larger values of voltage_c contribute negatively towards fault prediction. Overall, the beeswarm plot confirms that the classifier considers how the features respond among multiple electrical measurements rather than isolated feature thresholds.

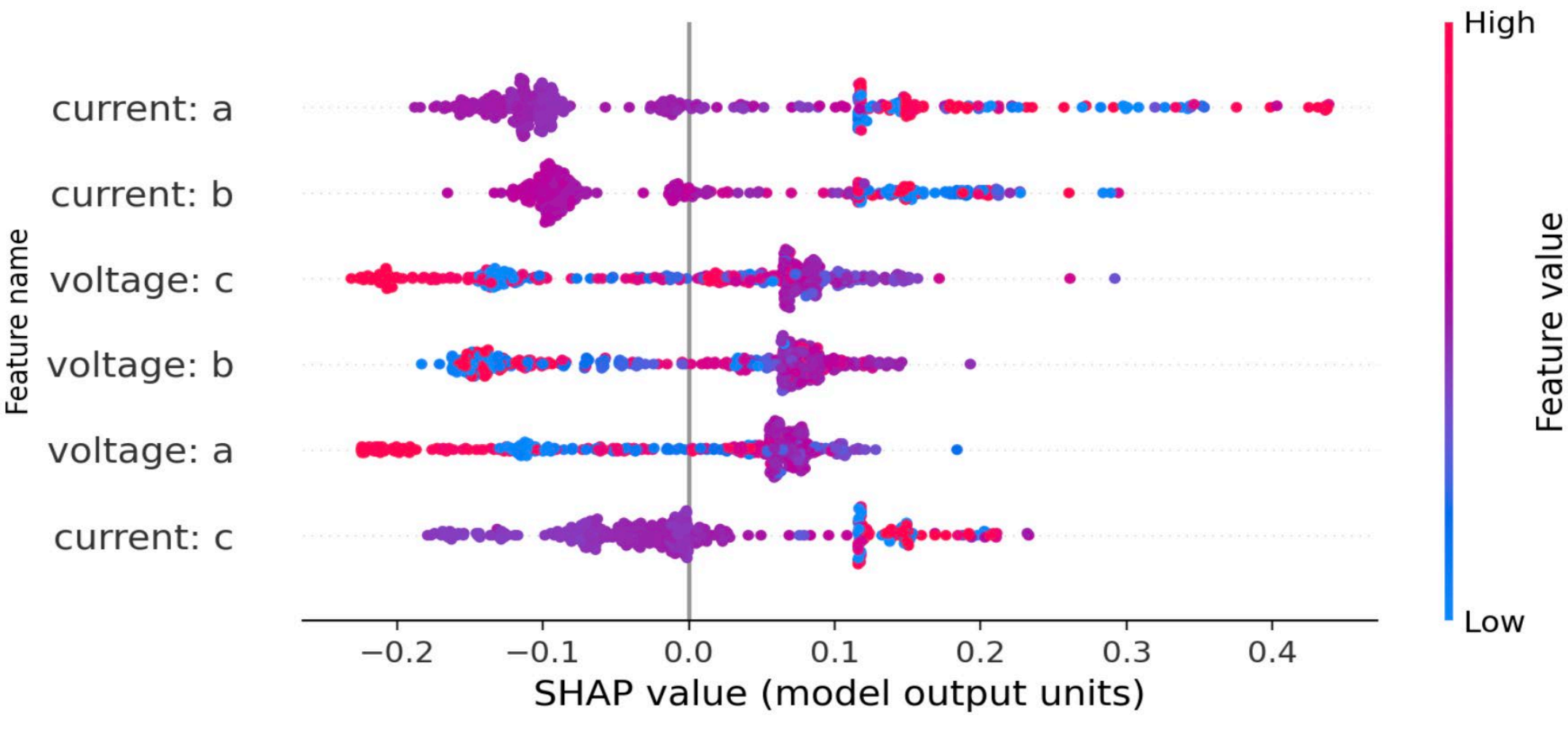

Fig 8: SHAP Summary plot

We also looked at SHAP explanations for two specific instances. For the instance that was correctly classified we started with the models baseline expected output of 0.445. The waterfall plot showed that voltage_c contributed -0.13, voltage_b contributed -0.13, current_a contributed -0.12 and current_b contributed -0.09. voltage_a had a positive effect of +0.05 while current_c had a small negative effect of -0.03. The combined effect of these contributions led to the prediction of 0, which correctly put the instance in the non-fault class. The voltage features voltage_c and voltage_b were the reasons for this correct decision, which makes sense based on what they physically mean.

For the instance that was misclassified the waterfall plot showed a different pattern. Starting from the baseline of 0.445 voltage_c contributed -0.21, voltage_b contributed +0.11, current_b contributed -0.11, current_a contributed -0.10, current_c contributed -0.07 and voltage_a contributed -0.07. Even though the contributions seemed to point to the fault class the model made an incorrect prediction. This shows that the models decision boundary is not always easy to understand, in complex cases. The conflicting contribution, from voltage_b combined with feature value combinations that the linear model did not capture likely led to this misclassification. This highlights a limitation of explanation methods: they may not always exactly reproduce the model's decision mechanism in complex cases.

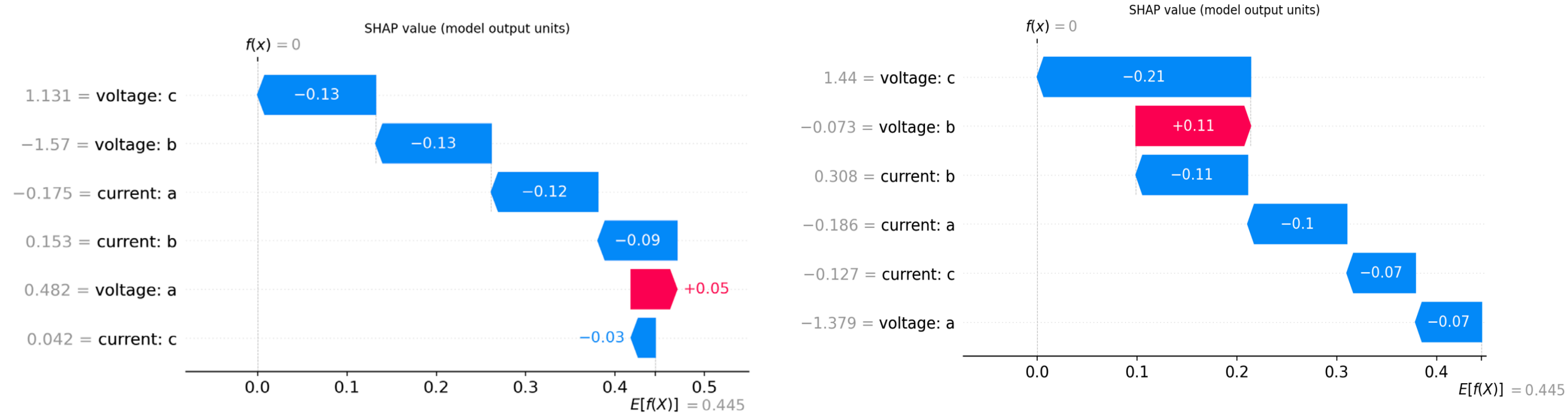

Fig 9: SHAP Explanations for 2 instances

### 4.1.4 Explanation Quality Metrics

The proposed framework was evaluated to see how well it explained the decisions it made. It was tested using two measures: Fidelity and stability, the explanations from the framework got a fidelity score of 1.00. This indicates SHAP explanations accurately reproduce the neural network's predictions without introducing approximation errors. With this score we can say explanations are faithful.

The stability score was also very high at 0.998. This shows that the explanations did not change much when the input was slightly different. This means the explanation method is not affected by errors in the data. So, the framework is very good at explaining things and its explanations are reliable when the input changes a bit. This is very important for using explanations in applications, like monitoring power systems, where things need to be safe. The proposed framework does a great job of explaining things in a way that is both accurate and reliable which is what we need for safety-critical power system monitoring applications.

## 4.2 Comparative analysis: neural network vs random forest – Fault Detection

When we compare the two models, the random forest model is better at predicting faults than the neural network. It gets a test accuracy and F1 score of 0.997, which is higher than the neural networks score of 0.99. Both models get a ROC-AUC score of 1.00. The random forest model also generalizes better with a gap between the training and test data.

In terms of explainability the two models are different. The random forest model gives importance to certain features like phase A and phase B currents. The neural network gives even importance to all features. However, both models agree that phase A and phase B currents are the important features. The random forest model also gives decisive local explanations for correctly classified instances. The neural networks explanations are a bit more diffuse, but still consistent.

The analysis of misclassifications shows that the random forest model can make mistake if the input data is extreme. The neural networks mistakes come from conflicts between features.

The neural network gets a fidelity score of 1.00 and a stability score of 0.998. The random forest model gets a fidelity score of 0.985 and a stability score of 0.987. The neural network have higher fidelity score which means that SHAP is better at reconstructing its predictions, whereas random forest model have lower fidelity score because it is harder to approximate a tree ensemble. However, both models have high fidelity and stability scores which means that the proposed framework gives reliable and high-quality explanations for both models. This makes it suitable, for safety-critical power system fault detection applications.

| **Aspect** | **Neural network** | **Random forest** |
|---|---|---|
| **Test Accuracy** | 0.990 | 0.997 |
| **Test F1-Score** | 0.990 | 0.997 |
| **ROC-AUC** | 1.000 | 1.000 |
| **SHAP Fidelity** | 1.000 | 0.985 |
| **SHAP Stability** | 0.998 | 0.987 |
| **Local Explanations (LIME)** | More diffuse but consistent explanations | More decisive and interpretable explanations |

Table 2: Comparison between NN & RF [Fault Detection]

### 4.3 Building energy regression

This part of the report shows what we found out when we used the XAI framework on the building energy labels dataset for predicting log-transformed site energy use (kBtu). The neural network regressor were assessed using predictive performance, LIME based explanations, SHAP based explanations, and explanation evaluation metrics. This analysis provides insight into both the performance and interpretability of the regression model.

### 4.3.1 Model Performance

The performance of the neural network regressor was evaluated using coefficient of determination ($R^2$), mean squared error (MSE) or (RMSE) on both the training and testing data. The model achieves training $R^2$ of 0.6884 and test $R^2$ of 0.6716, indicating 67% of variance in the log-transformed site energy consumption is explained by the learned non-linear relations. The corresponding training and test MSE values are 0.6188 and 0.6470 while RMSE values were 0.7867 ans 0.8040, respectively.

The close gap between training and testing performance tells us the model is generalizing well on unseen data reducing the risk of overfitting. Although the accuracy is lower than that of fault detection task, this behavior is expected because dataset is influenced by numerous factors. Many of these variables are either indirectly represented or unavailable within the dataset, making accurate energy prediction challenging.

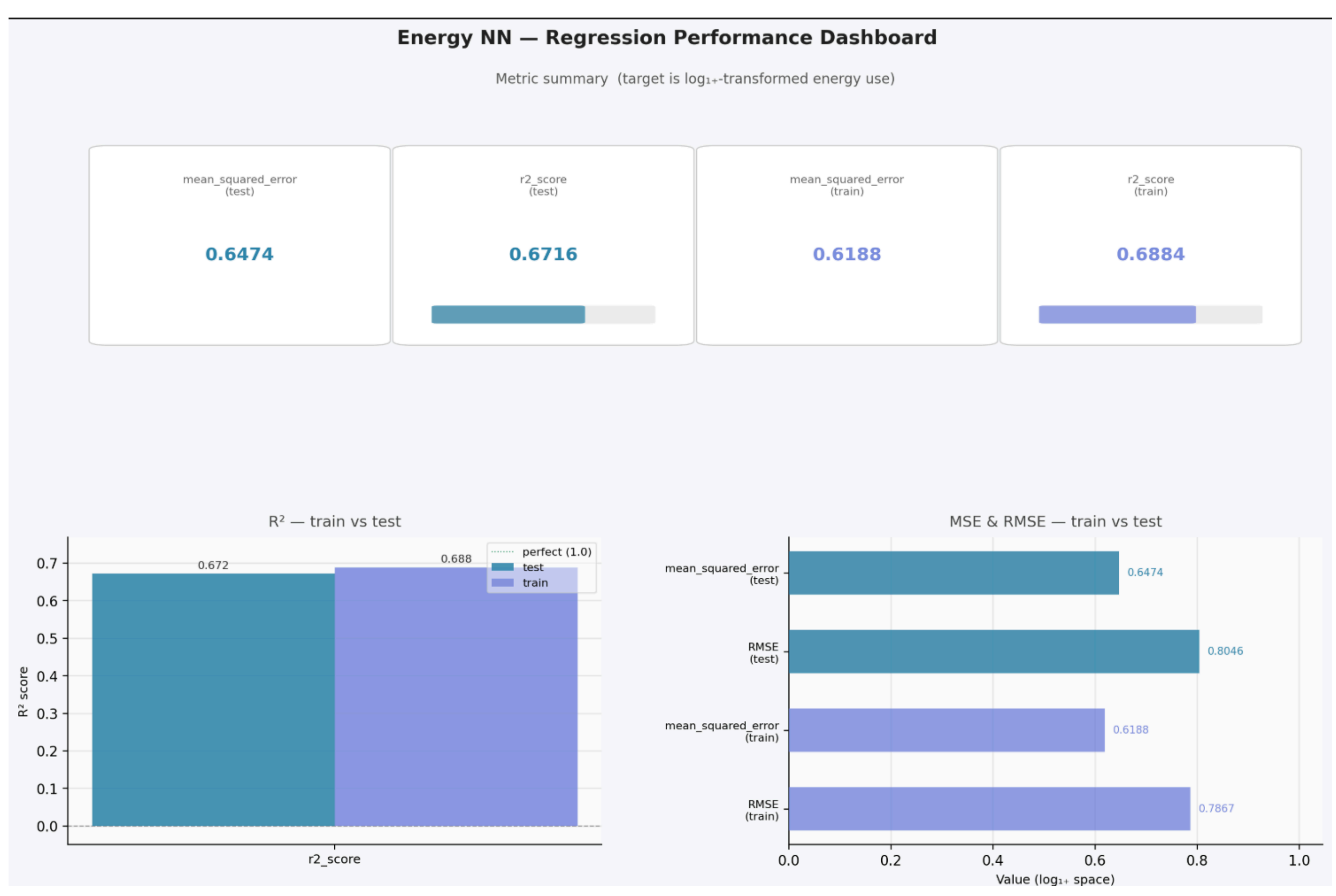

Fig 10: Energy regression performance metrics

### 4.3.2 LIME-Based local explanations

To interpret the neural network's predictions, LIME was applied to two representative test instances: one correctly predicted building and one incorrectly predicted building, as shown in figure below.

For correctly predicted instance building type and ProprtyGFATotal are dominant contributors to predicted energy consumption. Residential and Commercial buildings have largest positive contributions, while Institutional, Educational, NumberOfFLoors and Building age have comparatively smaller effects. This indicates that the model primarily relies on occupancy type and building size when estimating energy consumption.

For the misclassified instance a similar pattern of feature importance is observed with Commercial, Residential and PropertyGFATotal again contributing most strongly to the prediction. Despite these similar local explanations, the model produces an incorrect prediction, suggesting that the complex nonlinear relationships governing building energy consumption are not fully captured for this particular sample.

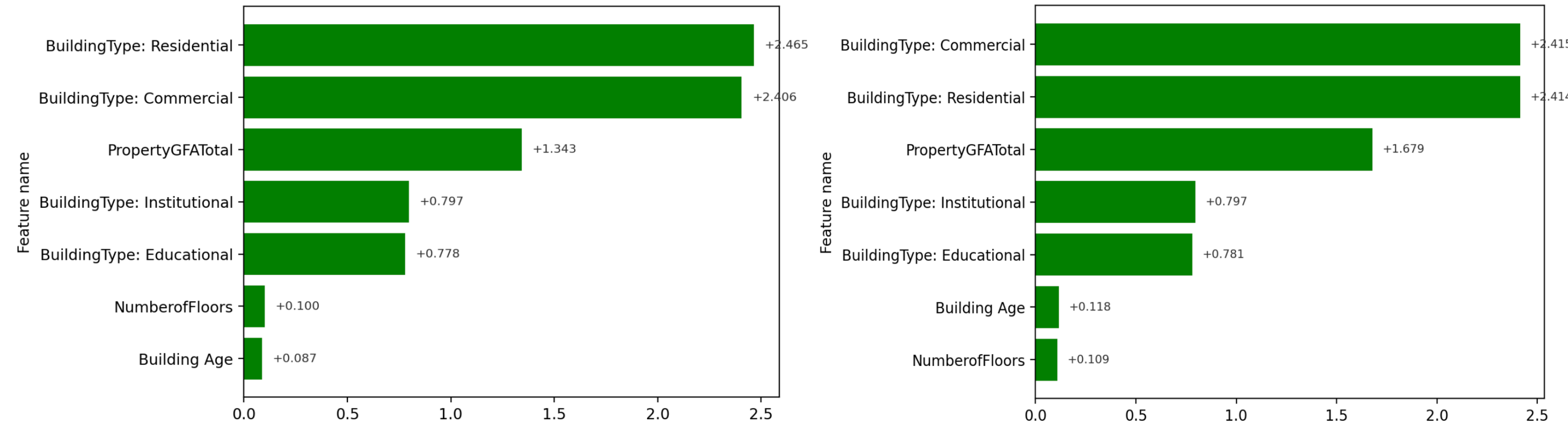

Fig 11: LIME explanations of 2 instances

**4.3.3 SHAP Based Global and Local Explanations**

The SHAP bar plot in the figure illustrates that building type is the most influential feature. The residential indicator is the most important feature with a mean absolute SHAP value of 2.40. The commercial indicator is right behind residential with a value of 2.38. These two features contribute the most in the decision making compared to other features, like PropertyGFATotal, which has a value of 0.51 Educational type with 0.34, NumberofFloors with 0.17, Building Age with 0.14 and institutional type with 0.10. This tells us that the neural network has figured out that building type is the crucial feature.

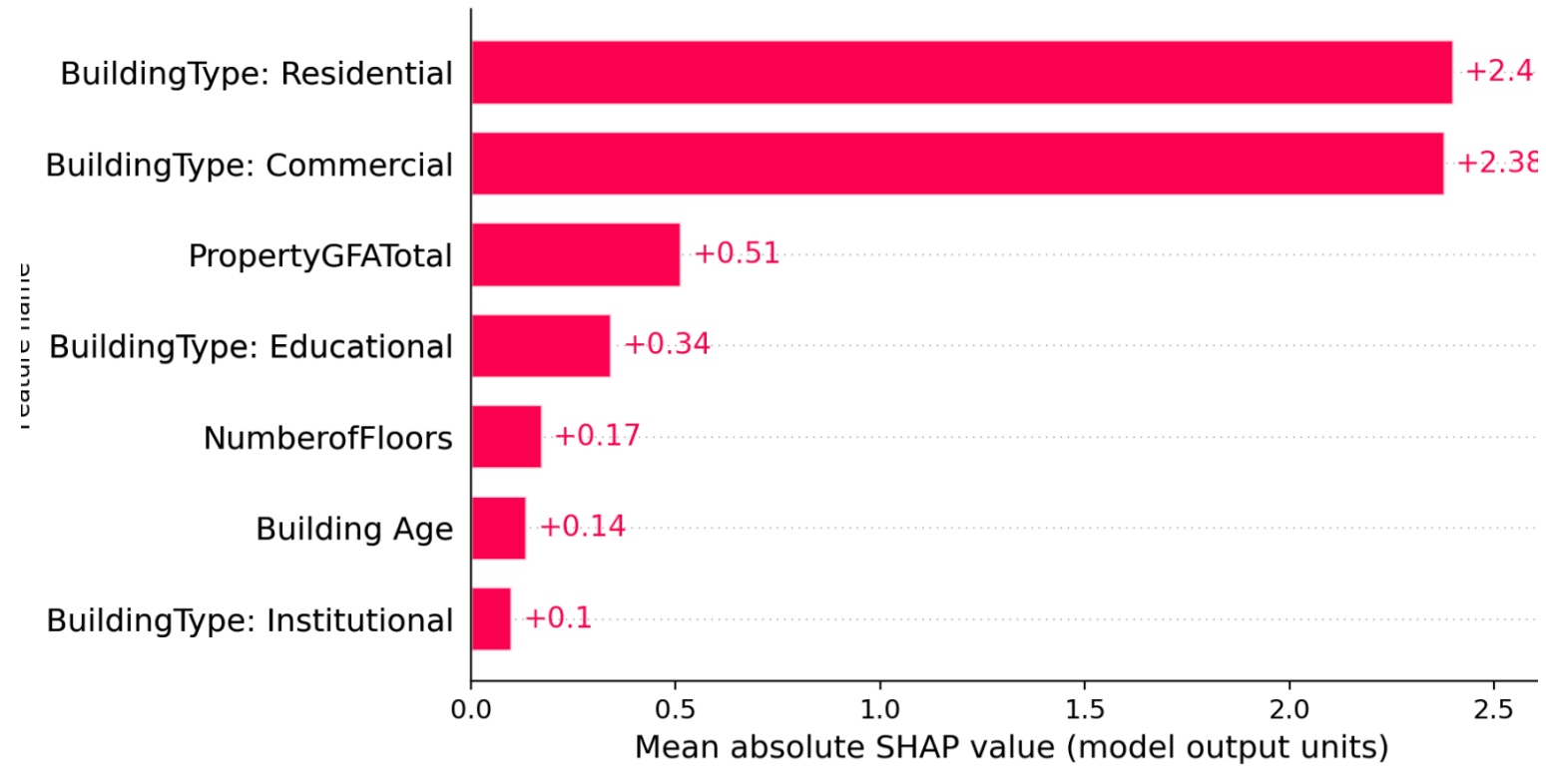

Fig 12: SHAP Bar plot

The SHAP beeswarm summary plot in figure below provides further insight into how individual feature values influence the predicted energy consumption across all samples. Residential and commercial building types exhibits largest SHAP value distributions, indicating that building type is the dominant factor influencing the neural network's predictions. Similarly, PropertyGFATotal shows an increasing positive contribution as building size increases, demonstrating that larger buildings are generally associated with higher energy consumption. On the other hand, Educational and Institutional building indicators exhibit comparatively smaller SHAP value distributions, suggesting that these features have a relatively limited influence on the model's predictions.

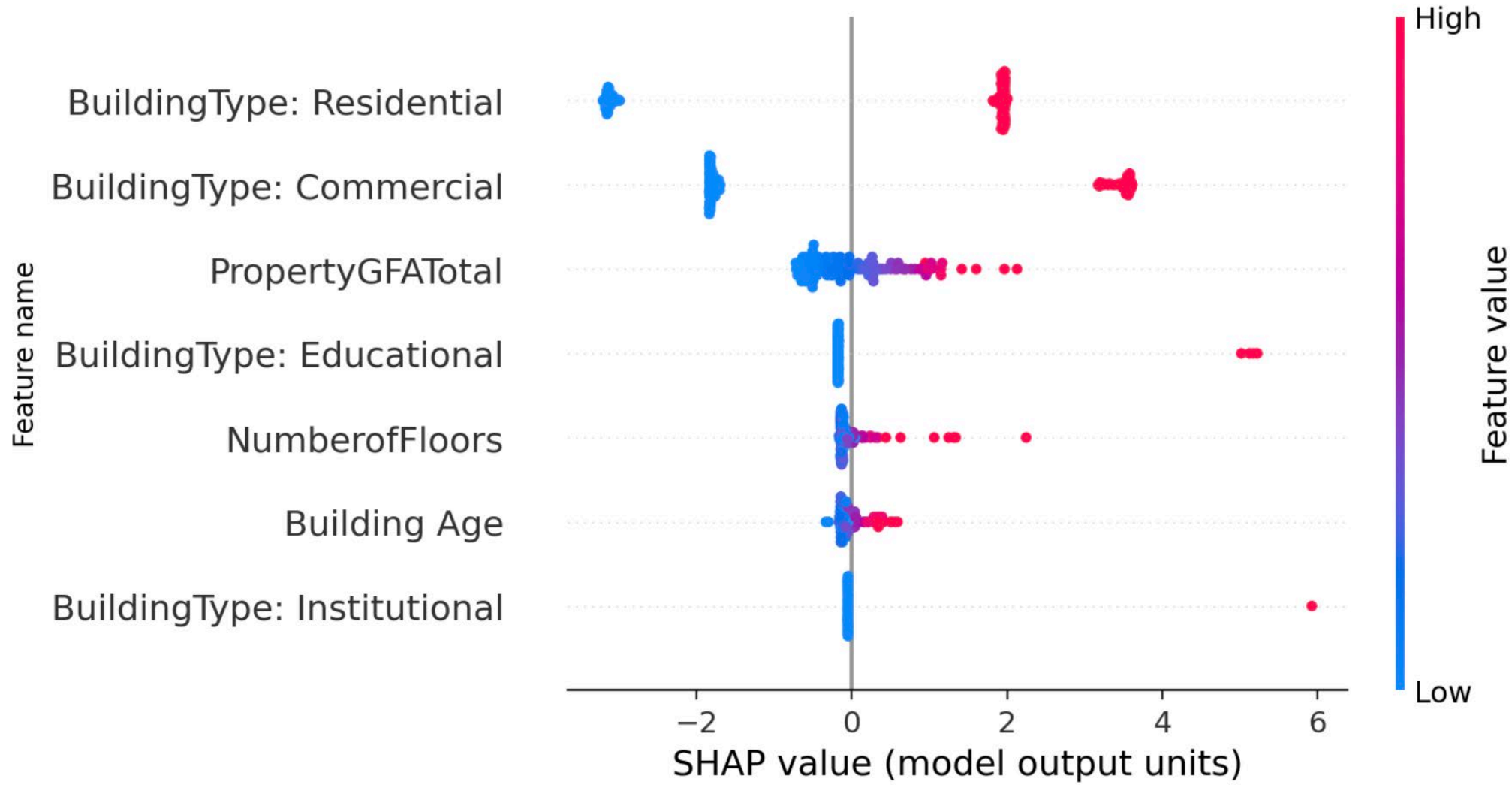

Fig 13: SHAP summary plot

To further investigate the model's decision-making process, SHAP waterfall plots were generated for one correctly predicted and one incorrectly predicted instance. For correctly predicted instance, the prediction is primarily driven by the positive contributions of the Residential building type and PropertyGFATotal, which outweigh the negative contribution of the commercial building type. On the other hand, for the misclassified instance, we can see strong contributions from commercial and residential but both are completely opposite to each other commercial is contributing positively residential is negatively, remaining all features have small effects. These feature contributions shift the prediction away from the true value, illustrating how complex interactions among building characteristics can lead to prediction errors.

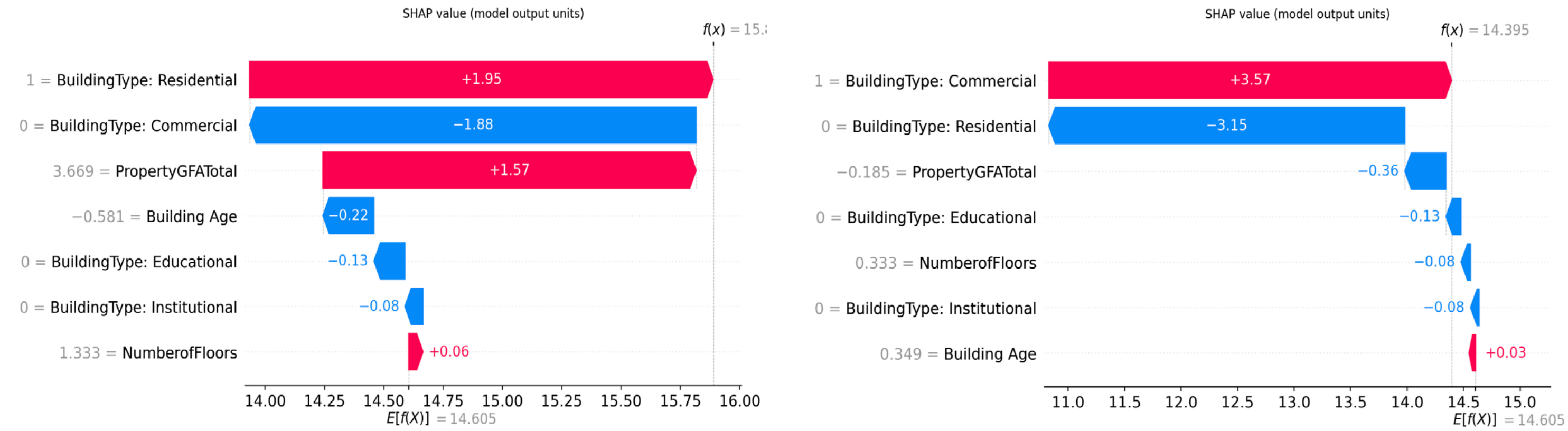

Fig 14: SHAP explanations for the 2 instances

### 4.3.4 Explanation Quality Metrics:

We use same evaluation metrics here as well i.e. stability and fidelity metrics. The model achieved fidelity score of 0.998, which indicates that the explanations closely approximate the model's predictions with minimal approximation error. When it comes to stability model achieved a score of 0.719 which is considerably lower than that obtained for the fault detection task, which tells us explanations are sensitive towards small changes in the input.

The reduced stability is expected because dataset consists of multiple characteristics. So, even small variations in highly influential features like building type or PropertyGFATotal can change the SHAP attribution which reflects in explanations. With these results we can say that neural network provides reliable energy predictions, the corresponding explanations are less robust than those of fault detection.

## 4.4 Comparative Analysis: Neural network vs random forest – energy regression

The two models are different when it comes to predicting energy use and explaining why they make particular predictions. This makes comparing them very important. When we look at how they predict energy use, the random forest model does better than the neural network model on the test set no matter what metric we use. For example, the random forest model gets a test $R^2$ of 0.7017 while the neural network model gets 0.6716. The random forest model also does better on test MSE and test RMSE.

However, the random forest model has a difference between its performance on the training set and the test set. This means it is more likely to overfit, which is not ideal. On the hand the neural network model generalizes more consistently but its performance is a bit lower.

The biggest difference between the two models is how they decide which features influence the prediction more. The neural network model thinks building type is the important factor, with residential and commercial buildings being the next top two factors. The random forest model on the other hand thinks the PropertyGFATotal of the building is the most important factor, with the type of building being less important. This shows that the two models have learned things from the same data. Both models are partially correct because energy use in buildings is affected by both the type of building and its size.

When it comes to explaining their predictions the random forest model is much better than the neural network model. The random forest model is very faithful to the model and very stable i.e. its explanations do not change much when the input data changes a bit. The neural network model is less stable which means its explanations can change a lot when the input data changes in the input data especially when it comes to categorical features like building type.

From a perspective the random forest model is more suitable, for real world applications because it provides more consistent and reliable explanations that engineers and policymakers can use to make decisions. The neural network model's explanations are less trustworthy because they can change a lot depending on the input data.

| Aspect | Neural network | Random forest |
| --- | --- | --- |
| **Test R2** | 0.6716 | 0.7017 |
| **Test MSE** | 0.647 | 0.617 |
| **Test RMSE** | 0.80 | 0.7855 |
| **SHAP Fidelity** | 0.998 | 0.999 |
| **SHAP Stability** | 0.719 | 0.982 |

Table 3: Comparison between NN and RF [Energy Regression]

## 5.Conclusion

This paper presented an explainable AI (XAI) framework for interpreting neural network models using LIME and SHAP, together with fidelity and stability metrics to quantitatively evaluate explanation quality. The framework was validated on two datasets: power systems fault detection and building energy consumption prediction, demonstrating its ability to provide both accurate predictions and meaningful explanations.

The results indicates the proposed framework explains the decision making of neural network at both local and global levels. For fault detection task explanation which were generated are highly stable and faithful. For the building energy regression task indicates building type and gross floor area are the most influential features, while lower stability score tells us about the complexity of the regression problem. Overall, the framework enables a more comprehensive evaluation of neural network models by combining predictive performance with explanation quality, thereby increasing transparency and confidence in their deployment for high-stakes engineering applications

Future work will focus on extending the framework to additional explainability techniques and evaluating its applicability to more complex deep learning models and diverse real-world datasets.